\documentclass[final,5p,times,twocolumn]{elsarticle}
\biboptions{sort&compress}
\pdfoutput=1

\usepackage{amssymb}
\usepackage{amsmath}
\usepackage{subfig}
\usepackage[figuresright]{rotating}

\usepackage{lineno}

\usepackage[labelfont=bf]{caption}
\usepackage[labelsep=period]{caption}

\graphicspath{{./}}

\journal{Phys. Lett. B}

\RequirePackage{xspace}
\def\Dbar {\kern 0.2em\overline{\kern -0.2em D}{}\xspace}
\def\DzDstarzb{\ensuremath{D^0\kern -0.16em \Dbar^{*0}}\xspace}

\newcommand{\upmumu}{\ensuremath{\Upsilon(1\text{S})\rightarrow\mu^{+}\mu^{-}}}
\newcommand{\Xb}{\ensuremath{X_{b}}}

\newcommand{\up}{\ensuremath{\Upsilon(1\text{S})}}
\newcommand{\upTw}{\ensuremath{\Upsilon(2\text{S})}}
\newcommand{\upTwupp}{\ensuremath{\Upsilon(2\text{S})\rightarrow\pi^{+}\pi^{-}\Upsilon(1\text{S})}}
\newcommand{\upTh}{\ensuremath{\Upsilon(3\text{S})}}
\newcommand{\upThupp}{\ensuremath{\Upsilon(3\text{S})\rightarrow\pi^{+}\pi^{-}\Upsilon(1\text{S})}}

\newcommand{\upFour}{\ensuremath{\Upsilon(4\text{S})}}

\newcommand{\upp}{\ensuremath{\pi^{+} \pi^{-} \Upsilon(1\text{S})}}

\newcommand{\upDJ}{\ensuremath{\Upsilon(1^3\text{D}_J)}}

\newcommand{\upDJTw}{\ensuremath{\Upsilon(1^3\text{D}_2)}}
\newcommand{\upDJTwupp}{\ensuremath{\Upsilon(1^3\text{D}_2)\rightarrow\pi^{+}\pi^{-}\Upsilon(1\text{S})}}

\newcommand{\invfb}{\ensuremath{\mathrm{fb}^{-1}}}
\newcommand{\kev}{\ensuremath{\mathrm{keV}}}
\newcommand{\mev}{\ensuremath{\mathrm{MeV}}}
\newcommand{\gev}{\ensuremath{\mathrm{GeV}}}
\newcommand{\tev}{\ensuremath{\mathrm{TeV}}}

\usepackage{preprintcover}
\PreprintCoverPaperTitle{Search for the \Xb\ and other hidden-beauty states in the \upp\ channel at ATLAS}
\PreprintIdNumber{CERN-PH-EP-2014-230}
\PreprintCoverAbstract{
This Letter presents a search for a hidden-beauty counterpart of the $X(3872)$
in the mass ranges 10.05--10.31 GeV and 10.40--11.00 GeV,
in the channel $X_b \to \pi^{+} \pi^{-} \Upsilon(1\rm S$$)(\rightarrow\mu^+\mu^-)$,
using 16.2 fb$^{-1}$ of $\sqrt{s}=8$ TeV $pp$ collision data
collected by the ATLAS detector at the LHC.
No evidence for new narrow states is found, and upper limits are set on the product
of the $X_b$ cross section and branching fraction,
relative to those of the $\Upsilon(2\rm S)$,
at the 95\% confidence level using the CL$_S$ approach. These limits
range from 0.8\% to 4.0\%, depending on mass.
For masses above 10.1 GeV,
the expected upper limits from this analysis are
the most restrictive to date.
Searches for production of the $\Upsilon(1^3\rm D$$_J)$, $\Upsilon(10860)$, and $\Upsilon(11020)$
states also reveal no significant signals.}
\PreprintJournalName{Physics Letters B}

\begin{document}

\begin{frontmatter}

\title{Search for the \Xb\ and other hidden-beauty states in the \upp\ channel at ATLAS}

\author{ATLAS Collaboration}

\begin{abstract}
This Letter presents a search for a hidden-beauty counterpart of the $X(3872)$
in the mass ranges 10.05--10.31 GeV and 10.40--11.00 GeV,
in the channel $X_b \to \pi^{+} \pi^{-} \Upsilon(1\text{S})(\rightarrow\mu^+\mu^-)$,
using 16.2 fb$^{-1}$ of $\sqrt{s}=8$ TeV $pp$ collision data
collected by the ATLAS detector at the LHC.
No evidence for new narrow states is found, and upper limits are set on the product
of the $X_b$ cross section and branching fraction,
relative to those of the $\Upsilon(2\text{S})$,
at the 95\% confidence level using the CL$_S$ approach. These limits
range from 0.8\% to 4.0\%, depending on mass.
For masses above 10.1 GeV,
the expected upper limits from this analysis are
the most restrictive to date.
Searches for production of the $\Upsilon(1^3\text{D}_J)$, $\Upsilon(10860)$, and $\Upsilon(11020)$
states also reveal no significant signals.
\end{abstract}


\end{frontmatter}

\section{Introduction}
\label{sec:introduction}

The $X(3872)$ is the first and the best-studied of the new hidden-charm
states seen in the last decade. Observed by Belle in decays
$B^{\pm}\to K^{\pm} X(\to\pi^+\pi^-J/\psi)$~\cite{belleX3872}, it was quickly confirmed
by BaBar~\cite{babarX3872}, CDF~\cite{cdfX3872}, and D\O~\cite{d0X3872}.
In particular, CDF and D\O\ found that the $X(3872)$ is produced directly in $p\overline{p}$
collisions; recently CMS has measured the product of the $pp$ production
cross section and the $\pi^+\pi^-J/\psi$ branching fraction
to be $(6.56\pm 0.29\pm 0.65)\%$ of the value for the $\psi(2S)$~\cite{cmsX3872}.
The mass, narrow width,
$J^{PC} = 1^{++}$ quantum number assignment~\cite{Abulencia:2005zc,Aubert:2006aj,Abulencia:2006ma,Choi:2011fc,lhcbX3872},
and decay characteristics of the $X(3872)$ make it unlikely to be a conventional
quarkonium state, and there is an extensive literature discussing its structure.
Weakly bound \DzDstarzb molecular models
(for example, Refs~\cite{tornqvist,swansonModel}) have been popular due to the
proximity of the $X(3872)$ to the \DzDstarzb threshold;
various $[qc][\bar{q}\bar{c}]$ tetraquark~(for example, Refs~\cite{maiani,relTetraquark})
and other models have also been proposed.

Heavy-quark symmetry suggests the existence of a hidden-beauty partner
--- a so-called \Xb\ state --- which should be produced in $pp$ collisions \cite{georgeXbpaper}.
The molecular model of Swanson~\cite{swansonModel,swanson}
predicts an \Xb\ mass of $10561\,\mev$, while tetraquark predictions vary:
for example, Ref.~\cite{relTetraquark} predicts masses of
10492, 10593, or $10682\,\mev$, depending on the flavour of the light quarks.

The decay $\Xb \to \pi^+ \pi^- \Upsilon(1S)(\to \mu^+\mu^-)$, analogous to the decay
mode in which the $X(3872)$ was discovered, provides a straightforward way to reconstruct
an \Xb. Any resulting measurement or upper limit on the \Xb\ production
cross section then depends on the branching fraction for
$\Xb \to \pi^+ \pi^- \Upsilon(1S)$, which is unknown.

The \upp\ channel also provides the opportunity to measure the production
of the \upDJ\ states. These have not been observed at the Tevatron;
their production cross sections in $pp$ collisions are also unknown,
but an early colour-octet calculation~\cite{up1Dproduction}
gives values comparable to that of the \upTw.
The \upDJTw\ has been observed in radiative transitions
by CLEO~\cite{cleo1D} and BaBar~\cite{babar1D}.

The production of $\Upsilon(10860)$ and some other hidden-beauty states may also
be studied using \upp.
The  $\Upsilon(10860)$ decay to \upp\ has a surprisingly large partial
width \cite{belleUp5Swidth,babarUp5Swidth,belleUp5SPartialWidth}: current world average results are $(0.29\pm0.15)\, \mev$ for the
$\Upsilon(10860)$, to be compared with $(0.89\pm0.08)\,\kev$ for the \upTh, and
$(1.7\pm0.2)\,\kev$ for the \upFour~\cite{pdg}.
Belle has also presented evidence
of exotic substructure in this decay~\cite{belle5Sdecays}.
The natural widths of the $\Upsilon(10860)$, $\Upsilon(11020)$,
and other states above the open-beauty threshold are larger than the detector
resolution, and must be explicitly considered in any search.

In 2013, CMS reported~\cite{cmsAnalysis} the results of their search for
$\Xb \to \pi^+\pi^-\up(\to\mu^+\mu^-)$,
finding no evidence for narrow states in the 10.06--$10.31\,\gev$ and
10.40--$10.99\,\gev$ mass ranges. They set upper limits on the product of
cross section and branching fraction at values between 0.9\% and 5.4\% of the \upTw\ rate.

This Letter presents a search for the \Xb\ and other hidden-beauty states
at ATLAS, using a $16.2\,\invfb$ $pp$ collision data sample collected at $\sqrt{s}=8\,\tev$
during the 2012 run of the LHC. The analysis is performed simultaneously
across eight kinematic bins of varying sensitivity; the \upTw\ and \upThupp\
peaks are used to validate the measurement technique.
Results are presented in terms of the product of production cross section and
\upp\ branching fraction, relative to that for the \upTw.

\section{The ATLAS detector}
The ATLAS detector \cite{detectorpaper} is composed of an inner tracking
system, calorimeters, and a muon spectrometer. The inner detector (ID)
surrounds the $pp$ interaction point and consists of silicon pixel and
microstrip detectors, and a transition radiation tracker,
all immersed in a 2 T axial magnetic field.
The ID spans the pseudorapidity\footnote{ATLAS uses a right-handed
	coordinate system with its origin at the nominal interaction point (IP)
	in the centre of the detector,
	the $x$-axis pointing to the centre of the LHC ring,
	and the $z$-axis along the beam pipe; the $y$-axis points upward.
	Cylindrical coordinates $(r,\phi)$
	are used in the transverse plane; $\phi$ is the azimuthal angle
	around the beam pipe. Pseudorapidity and transverse momentum
	are defined in terms of the polar angle $\theta$ as
	$\eta = -\ln(\tan{\theta/2})$ and $p_{\rm T} = p \sin{\theta}$.
	The $(\eta,\phi)$ distance between two particles is defined
	as $\Delta R = \sqrt{(\Delta \eta)^2 + (\Delta \phi)^2}$.
	For a particle with momentum $\vec{p}=(p_x,p_y,p_z)$ and energy $E$,
	the rapidity is defined as $y = 0.5\ln([E+p_z]/[E-p_z])$.}
range $|\eta| < 2.5$ and is enclosed by a system of electromagnetic
and hadronic calorimeters.
Surrounding the calorimeters is the muon spectrometer (MS) consisting of
three large air-core superconducting magnets (each with eight coils)
providing a toroidal field, a system of precision tracking chambers,
and fast detectors for triggering. Monitored drift tubes and cathode-strip
chambers provide precision measurements in the bending plane of muons
within the pseudorapidity range
$|\eta| < 2.7$.
Resistive plate and thin gap chambers
are used to make fast event data-recording decisions in the ranges
$|\eta|<1.05$ and $1.05 < |\eta| < 2.4$ respectively,
and also provide position measurements in the non-bending plane and improve
pattern recognition and track reconstruction. MS momentum measurements
are based on track segments formed in at least two of the three
precision chamber planes.

The ATLAS detector employs a three-level trigger~\cite{atlasTrigger}
to reduce the 20 MHz proton bunch collision rate to the few-hundred hertz
transfer rate to mass storage. This analysis is based on a Level-1 muon trigger that searches for hit
coincidences between muon trigger detector layers inside
pre-programmed geometrical windows that bound the path of muon candidates
above a given $p_{\rm T}^\mu$ threshold, and provide a rough estimate
of their position, for $|\eta^\mu| < 2.4$.
There are two subsequent software-based trigger stages,
in which muon candidates incorporate, with increasing precision,
information from both the MS and the ID, reaching position and momentum
resolution close to that provided by the offline reconstruction.

\section{Reconstruction and event selection}
\label{sec:reconEventSel}

Events are selected using a trigger requiring two muons of opposite charge,
each with $p_{\rm T}^\mu>4\,\gev$,
successfully fitted to a common vertex.
The $\mu^+\mu^-$ mass range accepted by the trigger, 8--$12\,\gev$, includes the \up, \upTw, and \upTh\ signal peaks.

In the offline reconstruction for this analysis,
muon reconstruction relies on a statistical combination of
an MS track and an ID track.
The selected muons are restricted to $|\eta^\mu| < 2.3$,
ensuring high-quality tracking and a reduction of fake muon candidates.
This restriction also removes regions of strongly varying efficiency and acceptance.

The reconstruction of \upp\ candidates begins with pairs of
oppositely charged muon candidates
that satisfy the same kinematic conditions used by the trigger,
and have $\geq\!2$ pixel and $\geq\!6$ silicon microstrip detector hits.
Each pair is subjected to a common vertex fit, and a loose chi-square selection
is imposed to exclude very poor candidates.
Any dimuon with an invariant mass within $350\,\mev$ of the \up\ mass~\cite{pdg},
$m_{\text{1S}}$, is retained and considered
an \upmumu\ candidate. To confirm that this pair is the same as that used
in the trigger, the reconstructed and trigger-level muons are required to
match with $\Delta R < 0.01$.

In the remainder of the event, dipion candidates are formed from
oppositely charged pions with $|\eta^\pi|<2.5$, each required to have $\geq\!1$ pixel hits, $\geq\!6$
silicon microstrip hits, and $p_{\rm T}^\pi > 400\,\mev$;
no other requirements (such as lepton vetoes) are imposed.
The \up\ candidate and the dipion system are combined by
performing a four-track common-vertex fit, with the $\mu^+\mu^-$ mass
constrained to $m_{\text{1S}}$, and tracks assigned $\mu$ or $\pi$ masses
as appropriate. This significantly improves the mass resolution:
for example, in the \upTw\ simulation, the RMS improves from 142 MeV to 9.7 MeV.
The \upp\ vertex fit is required to
have a chi-square less than 20; this is 95$\%$ efficient for \upTw\ decays
but reduces background by a factor of ${\sim} 10$.
All remaining \upp\ candidates with
invariant masses up to $11.2\,\gev$ are retained.

For a state decaying to \upp, the acceptance $\mathcal{A}$ is defined
as the fraction of decays where both muons have $p_{\rm T}^\mu>4\,\gev$, both pions have
$p_{\rm T}^\pi>400\,\mev$, and all four particles are within $|\eta|<2.5$.
States with $p_{\rm T}<5\, \gev$ or rapidity $|y|>2.4$ have very low
acceptance, so candidates in these regions (which also suffer from high
background) are excluded.

\section{Data and simulation samples}
\label{sec:dataSimSamples}
The techniques adopted in this analysis were developed using $\sqrt{s}=7\,\tev$
data collected in 2011 and simulation samples generated under the same
running conditions, with particular attention to \upTwupp\ mass and
$(|y|,p_{\rm T})$ distributions. Backgrounds due to inclusive \up\ production
and combinatorial $\mu^+\mu^-$ sources were studied using
$\mu^+\mu^-$ sideband and $\mu^{\pm}\mu^{\pm}$
same-sign samples, and found to be featureless
above $9.8\,\gev$.

The results presented here are based on data from the 2012 $\sqrt{s}=8\,\tev$
$pp$ run at the LHC; standard data-quality criteria are used to ensure
efficient detector performance. The trigger employed in the event
selection was subject to an instantaneous luminosity-dependent prescale factor,\footnote{
An un-prescaled trigger with a $p_{\rm T}^\mu=6$ GeV threshold on the muons was also considered,
but found to result in a lower sensitivity using the optimisation procedure described in Section~\ref{sec:fitting}.} and provided an
integrated luminosity of $\mathcal{L}=16.2\,\invfb$. The resulting data sample includes over
10 million $\mu^+\mu^-$ combinations: a fit to the sample finds $(6.00\pm0.01) \times 10^6$ from \up\ decays,
$(0.200\pm0.002)\times10^6$ from \upTw\ decays, and the remainder from
combinatorial background. Given the reconstruction and event selection choices
described in the previous section, each dimuon gives rise (on average) to
19.5 \upp\ candidates with invariant mass below $11.2\,\gev$.
A procedure to select at most one candidate per event was considered.
However, after the adoption of the binning approach described in Section~\ref{sec:fitting},
candidate selection was found to worsen the expected sensitivity to a hypothetical \Xb\ signal.
Therefore, all \upp\ candidates are retained for analysis.

Simulated samples are used to optimise the selections, model signal decays,
develop fitting models, and calculate efficiencies.
Individual samples are used for the \upTw, \upTh, \upDJ\ triplet,
$\Upsilon(10860)$, and for two hypothetical \Xb\ masses, $10233\,\mev$ and $10561\,\mev$.
Production is modelled with
the AU2~\cite{pythia8tunes} tune of \textsc{Pythia} 8.170~\cite{pythia8} and the CTEQ6L1~\cite{cteq6l1} parton distribution functions.
In the decay to \upp, the three-body phase space is uniformly sampled.
Isotropic spin alignment of the parent is assumed:
for \upTw\ and \upTh\ this is supported by
a recent CMS measurement~\cite{cmsUpsilonPolarisation}.
Passage of particles
through the detector is simulated with Atlfast II~\cite{atlfastII},
supplementing \textsc{Geant4} \cite{geant4original,geant4update} with a parameterised calorimeter response.

Simulated kinematic distributions of the final-state pions and muons are
sensitive to mismodelling of the $p_{\rm T}$ and $y$ distributions of the parent
state.
ATLAS has measured doubly differential production cross sections for \upTw\
and \upTh\ at $7\,\tev$
in $p_{\rm T}$ bins up to $70\,\gev$
for $|y|<1.2$ and $1.2<|y|<2.25$~\cite{upProdXsec}.
These results can be extended to $|y|<2.4$ assuming flat rapidity dependence,
and up to $p_{\rm T}=100\,\gev$ using CMS measurements~\cite{upProdXsecCMSNewest}.
The resulting \upTw\ cross section is compared
to the production kinematics of a $7\,\tev$ simulation
using the 2011 ATLAS tune~\cite{mc11tune} of \textsc{Pythia} 6.4~\cite{pythia6}.
The ratio of the two in $(|y|,p_{\rm T})$ bins defines \textit{production weights},
which are applied to the $8\,\tev$ \upTw\ simulated sample, assuming that
\textsc{Pythia} correctly models the increase in cross section with $\sqrt{s}$.
The same procedure is used for the \upTh; for other masses,
linear extrapolation of the \upTw\ and \upTh\ weights is used. The simulated
\upTw\ and \upThupp\ samples are further reweighted to match dipion
mass distributions observed by CLEO~\cite{cleo2Sto1Spipi,cleoUpTransitions},
to allow comparison of simulation and data.

\section{Fitting strategy}
\label{sec:fitting}

Due to the low pion momentum in the \upp\ rest frame,
and the reconstruction threshold of $p_{\rm T}^\pi=400\,\mev$ in the laboratory
frame, true \upp\ decays are preferentially reconstructed if the parent has
large $p_{\rm T}$ or small $\theta^*$ (defined as the angle, in the parent rest frame,
between the dipion momentum and the lab-frame parent momentum).
In background candidates, the dipion and dimuon systems are unrelated, and the di\-pions typically
have low $p_{\rm T}^{\pi\pi}$ in the lab;
boosting to the $\pi^+\pi^-\mu^+\mu^-$ frame yields
large values of $\theta^*$,
with a broad distribution around $\cos{\theta^{*}}=0$.
In the $(p_{\rm T}, \cos{\theta^{*}})$ plane, then,
the ratio of signal to background candidates is largest in the upper-right
region, and smallest in the lower-left.

The mass resolution and background shape differ at central and
forward rapidities, so the analysis is performed in bins of
$|y| < 1.2$ (barrel) and $1.2 < |y| < 2.4$ (endcap).
Several possible ways to exploit the $(p_{\rm T}, \cos{\theta^{*}})$ discrimination
were considered, including further binning,
a diagonal cut in $(p_{\rm T}, \cos{\theta^{*}})$, and
a requirement on the $\Delta R$ between the \up\ and each pion
(as used by CMS~\cite{cmsAnalysis}).
The choice of method was based on optimising the expected significance for
a weak signal at a mass of $10561\,\mev$.
In the final approach eight analysis bins of varying sensitivity are used,
formed from combinations of high and low $|y|$, high and low $p_{\rm T}$,
and high and low $\cos{\theta^{*}}$. The optimal bin boundaries for $p_{\rm T}$
and $\cos{\theta^{*}}$ were determined to be $20\,\gev$ and 0, respectively.

\begin{figure}
  \centering
  \includegraphics[width=0.48\textwidth]{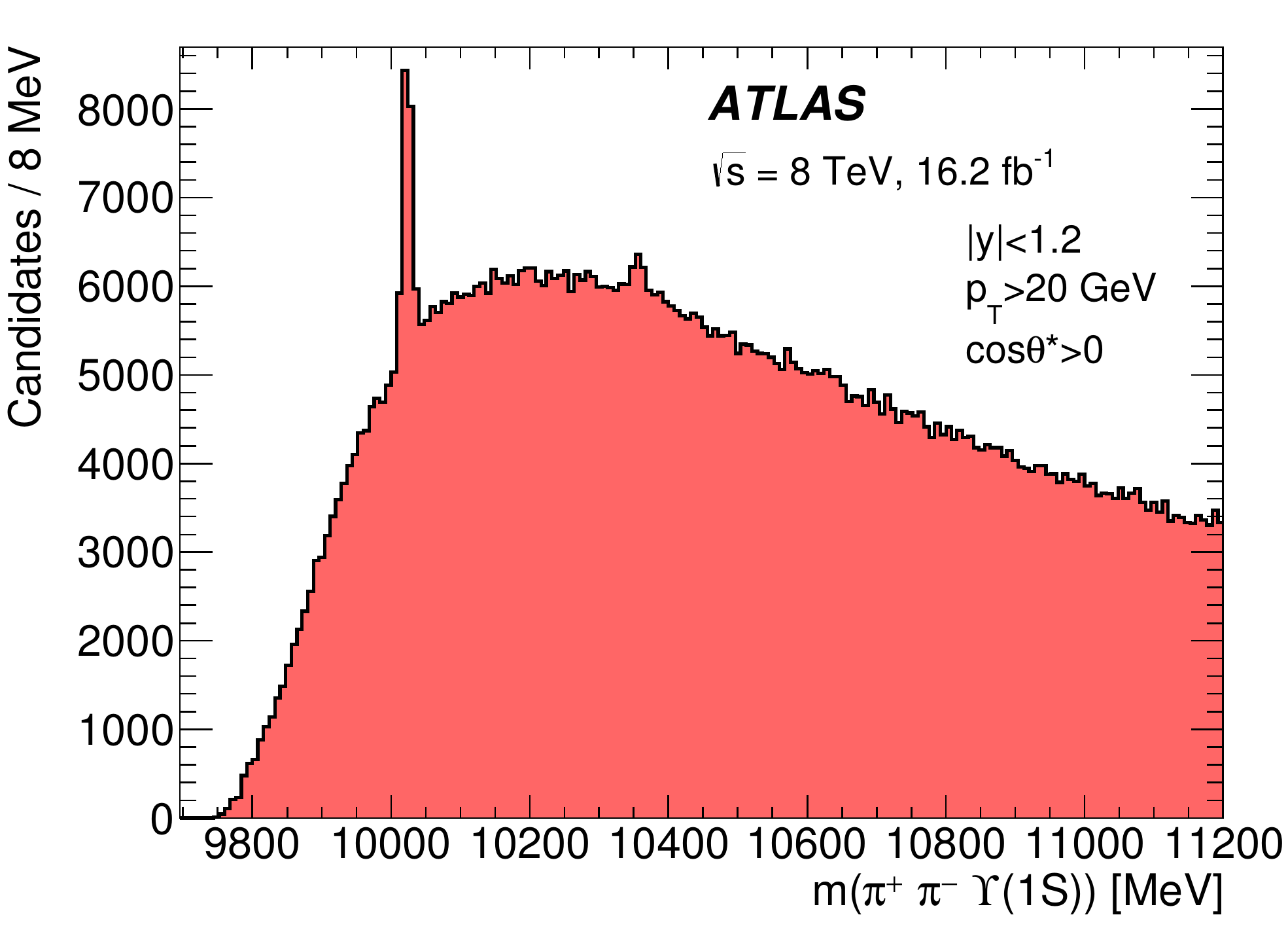}
  \caption{The \upp\ invariant mass distribution in the kinematic bin most sensitive to an \Xb\ signal:
	$|y|<1.2$, $p_{\rm T}>20$ GeV, and $\cos{\theta^{*}}>0$.
	The only apparent peaks are at the masses of the \upTw\ ($10023\,\mev$)
	and \upTh\ ($10355\,\mev$).}
  \label{fig:upPiPiData2012_yL_ptH_ctsH_inPaper}
\end{figure}

The \upp\ mass distribution for the most sensitive bin
($|y|<1.2$, $p_{\rm T}>20\,\gev$, $\cos{\theta^{*}}>0$)
is shown in Fig. \ref{fig:upPiPiData2012_yL_ptH_ctsH_inPaper}.
The \upTw\ and \upTh\ peaks are clearly visible, but no other peaks are apparent.
The background in this bin decreases with mass above the \upTh, whereas the fraction of signal events falling in this bin is constant above the \upTh\ in the simulation. Thus higher sensitivity is expected at larger masses.

Based on the simulation samples, the fraction of signal in the barrel region
$|y|<1.2$ is independent of mass with an average value of
$S_{|y|}=0.606\pm0.004$. Within the barrel, the fraction of signal with
$p_{\rm T}<20\, \gev$ develops smoothly with mass and can be characterised
by an analytic turn-on curve $S^{\rm b}_{p_{\rm T}}(m)=a/(1+e^{-b(m-c)})$.
In the endcap, the dependence is described by $S^{\rm ec}_{p_{\rm T}}(m)$,
which has the same functional form as $S^{\rm b}_{p_{\rm T}}(m)$
with different values for the parameters.
Similarly, within each $(|y|, p_{\rm T})$ bin the fraction of the signal
with $\cos{\theta^{*}}<0$ is modelled with a quadratic function,
$S^{\rm (i)}_{\cos{\theta^{*}}}(m)=a+bm+cm^2$, where $i=1$--$4$ labels the bin.
These $S$ functions, seven in total, are referred to below
as \textit{splitting functions}.
At any specified mass, the signal yield fraction in any particular
$(|y|,p_{\rm T},\cos{\theta^{*}})$ bin can be calculated from an
appropriately chosen product of three of these and their complements, $(1-S)$.
For example, the fraction in the bin
$(|y|<1.2,p_{\rm T}>20\, \gev,\cos{\theta^{*}}<0)$ is given by
$S_{|y|}\cdot(1-S^{\rm b}_{p_{\rm T}}(m))\cdot S^{(3)}_{\cos{\theta^{*}}}(m)$.

The shape of the signal peaks reflects the detector resolution,
which differs between the barrel and endcap, and varies as a function of mass;
in a given rapidity bin at a given mass, a single function can be used across the whole
$(p_{\rm T}, \cos{\theta^{*}})$ range. In each rapidity bin, the signal is fitted
using two Gaussians with a common mean, a narrow component fraction $f$,
and a ratio $r$ of broad to narrow widths. These parameters are found to be
independent of mass, and are fixed to the average values across the simulation
samples. The remaining parameter, the width of the narrow component, $\sigma$,
depends linearly on the mass of the parent state. Together with the splitting
functions defined above, this allows the signal shape and the fraction of the
signal falling in each of the analysis bins to be determined for any \Xb\ mass.

Searches for the production of the $\Upsilon(10860)$ and $\Upsilon(11020)$,
which have natural widths larger than the experimental resolution, are
also performed. Each of these states is modelled as a
Breit--Wigner convolved with the mass-dependent signal shape described in the previous paragraph,
representing the detector resolution. The fractions of the $\Upsilon(10860)$
signal falling in the eight analysis bins are extracted from the simulation,
while those for the $\Upsilon(11020)$ are determined by extrapolation.

All fits presented here are binned, extended maximum-likelihood fits in a local
region around the mass of interest. The bin width is $2\,\mev$ in all fits.
The background is described by a
linear combination of Chebychev polynomials up to second-order,
with independent parameters in each analysis bin,
unless otherwise specified.

\section{Results for the \upTw\ and \upTh}
\label{sec:up2S3Sresults}
\begin{figure*}[t]
  \centering
  \subfloat[\upTw: $|y|<1.2$]{\label{fig:upPiPiData2012_up2SFree_barrel_fitted}\includegraphics[width=0.33\textwidth]{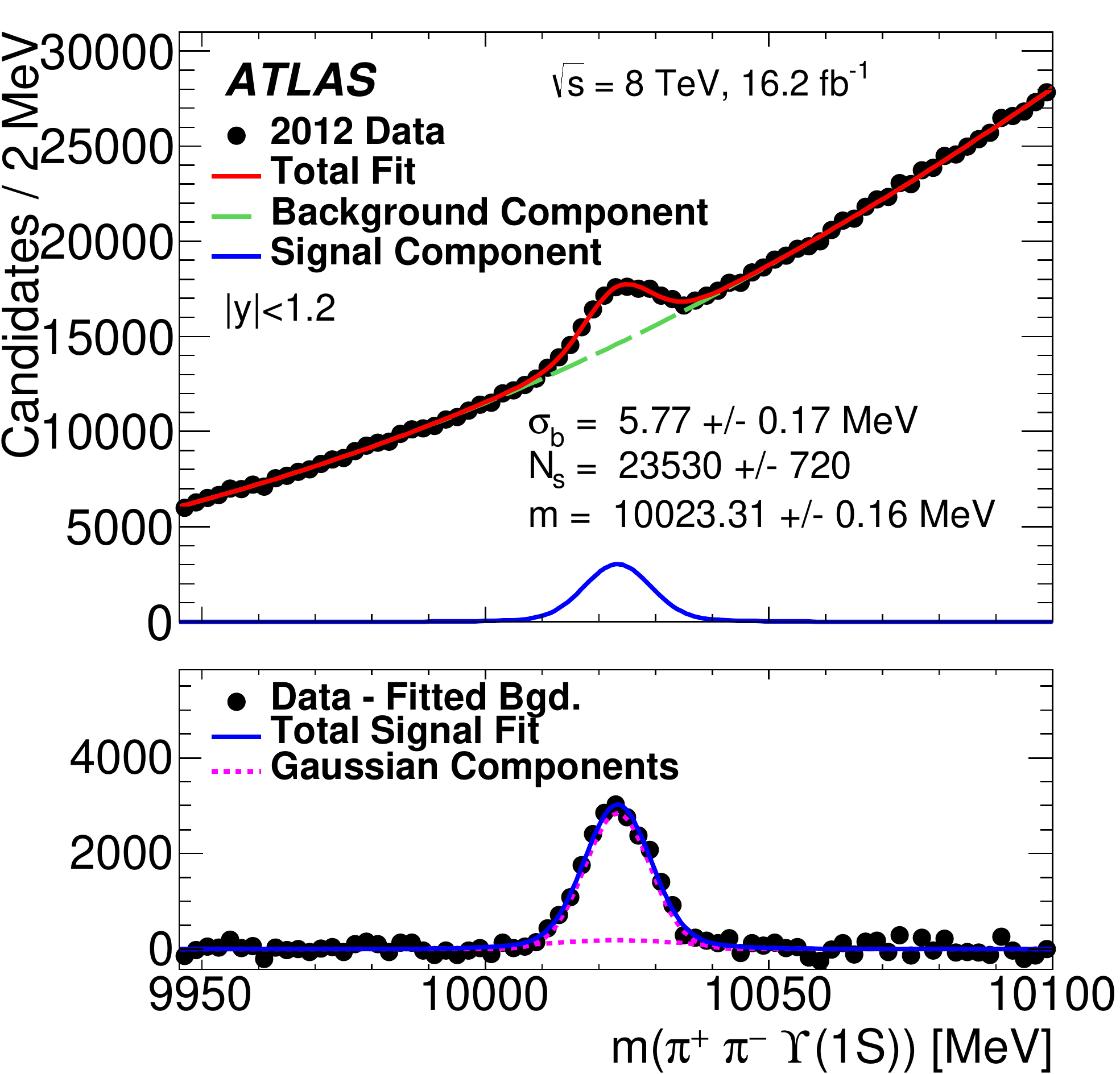}}%
  \subfloat[\upTw: $1.2<|y|<2.4$]{\label{fig:upPiPiData2012_up2SFree_endcap_fitted}\includegraphics[width=0.33\textwidth]{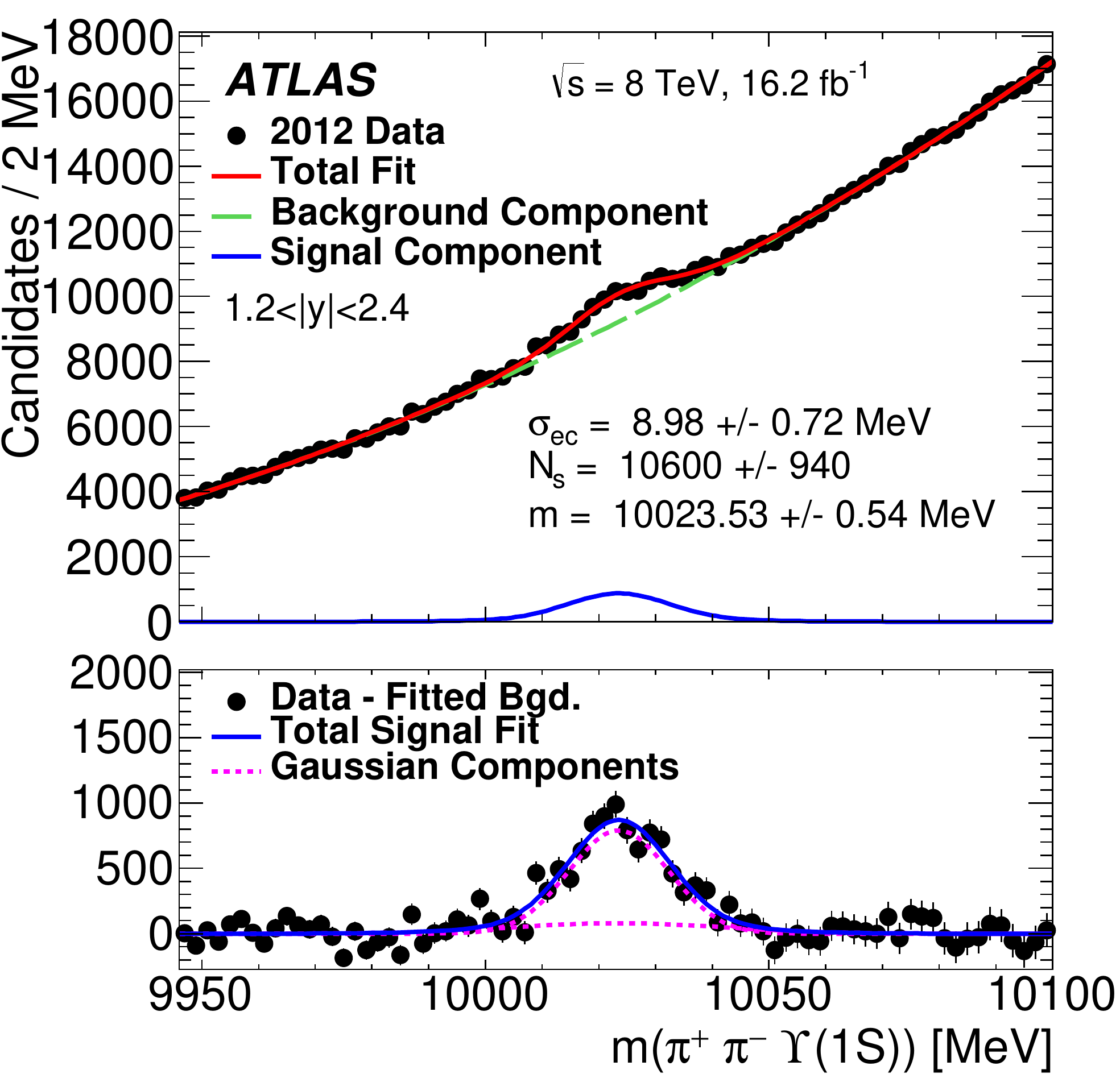}}%
  \subfloat[\upTh: $|y|<1.2$, $p_{\rm T}>20\,\gev$, $\cos{\theta^*}>0$]
	{\includegraphics[width=0.33\textwidth]{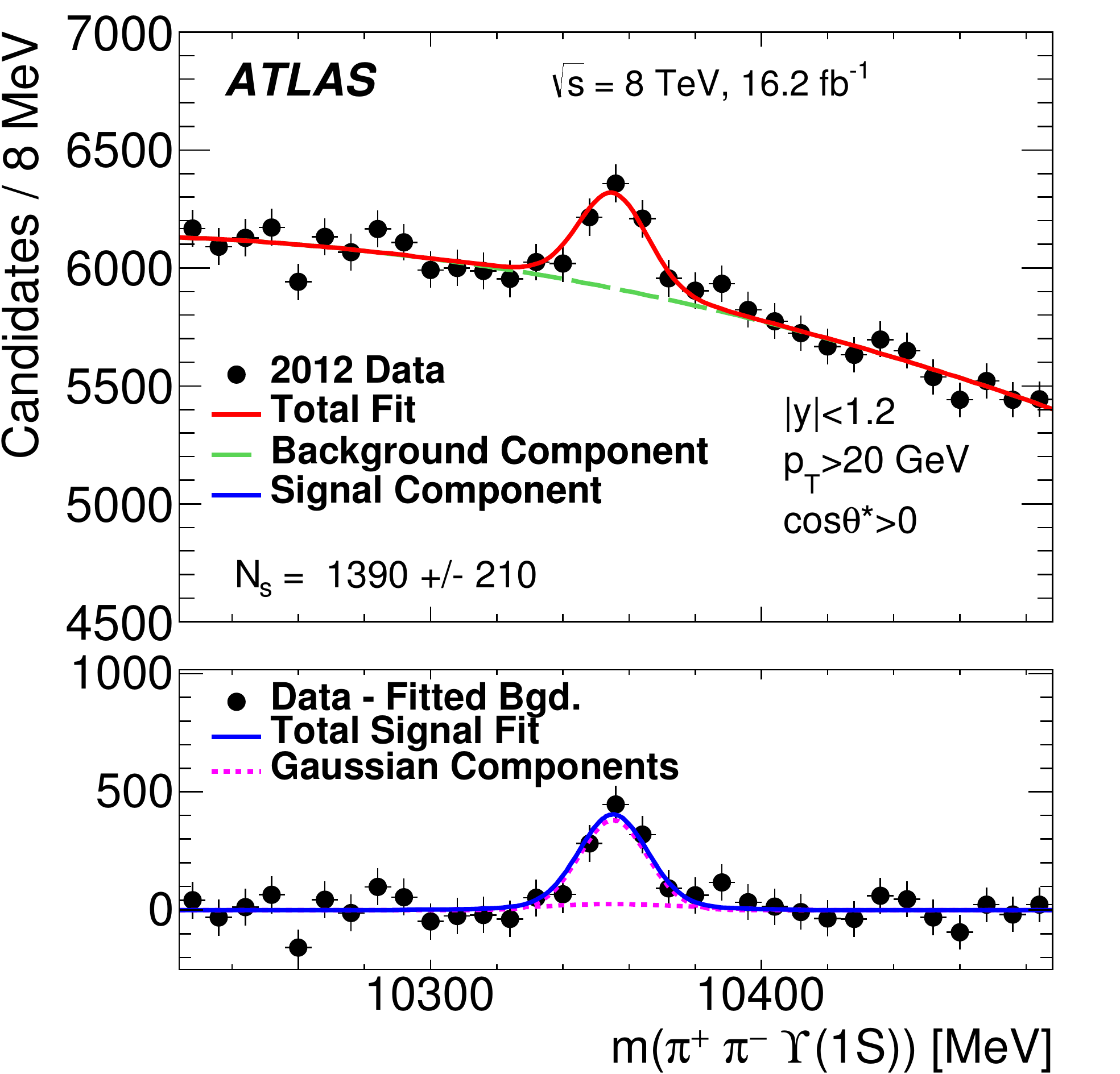}}
  \caption{The measured \upp\ invariant mass (data points),
	together with fits (red solid curves)
	to the \upTw\ peak in the (a) barrel and (b) endcap,
	with signal mass ($m$) and width parameters ($\sigma_{\text{b}}$ and $\sigma_{\text{ec}}$, respectively) free; and
	(c) a fit to the \upTh\ peak in the most sensitive bin,
	with zero suppressed on the vertical scale. $N_{\text{s}}$ is the fitted signal yield
        in each bin.
        The background component (green, long-dashed) and the signal component
	(peaked blue curve) are also shown separately.
	In each case, the lower panel shows background-subtracted data,
        with the total signal function (blue solid) and its two Gaussian components (pink, short-dashed curves) overlaid.}
  \label{fig:upPiPiData2012_up2SFree_fitted}
\end{figure*}

Fits to the \upp\ spectrum near the \upTw\
are first performed separately in barrel and endcap bins,
across the full $(p_{\rm T}, \cos{\theta^{*}})$ range,
with signal mass and width parameters free in the fit
(Figs~\ref{fig:upPiPiData2012_up2SFree_barrel_fitted} and \ref{fig:upPiPiData2012_up2SFree_endcap_fitted}).
In both cases, the mass is consistent with the world average for the \upTw,
and the $\sigma$ parameters are within uncertainties of the values
fitted to the \upTw\ simulation.

Signal shape parameters are then fixed to simulated values to reduce
uncertainties, and a separate fit is performed on each analysis bin.
The fraction of signal decays falling in the barrel is measured to be
$0.67\pm0.04$, compared to $0.606\pm0.004$ in the simulation;
similar results were seen in $7\,\tev$ data.
The ratio $0.67/0.606$ is used to rescale the barrel fraction
in the simulation.

The total fitted yield $N_{\text{2S}}=34300\pm800$ is consistent with
\begin{equation}
\label{eqn:yieldFormula}
N_{\text{2S}}^{\text{expected}} = (\sigma\mathcal{B})_{\text{2S}} \cdot \mathcal{L} \cdot \mathcal{A} \cdot \epsilon
 = 33300\pm2500,
\end{equation}
where the product of the cross section and branching fraction, $(\sigma\mathcal{B})_{\text{2S}}$, is estimated from the extended cross-section
measurement (see Section~\ref{sec:dataSimSamples}) using world-average values for the branching fractions
$\mathcal{B}(\up \rightarrow \mu^+\mu^-)$ and $\mathcal{B}(\upTwupp)$~\cite{pdg}.
The shape of the doubly differential cross section is also used to calculate
the acceptance, $\mathcal{A}=(1.442\pm0.004)\%$,
assuming the CLEO dipion mass spectrum~\cite{cleo2Sto1Spipi} and isotropic signal decays.
The reconstruction efficiency for decays within the acceptance,
$\epsilon=0.283\pm0.002$, is taken from the \upTw\ simulation.
To test kinematic distributions in the \upTw\ simulation,
signal fractions in the eight analysis bins are checked against
their expected values, and are found to be consistent within
statistical uncertainties.

A simultaneous fit to the analysis bins is also performed at the \upTh\ mass,
with reduced $\chi^2 = 1.0$ and statistical
significance (from the likelihood-ratio test statistic) $z=8.7$.
An individual fit to the most sensitive bin
(shown in Fig.~\ref{fig:upPiPiData2012_up2SFree_fitted}(c), with
larger binning to emphasise the peak)
has significance $z=6.5$.
The total \upTh\ yield,
$11600\pm1300$, agrees with the prediction estimated analogously to Eq.~(\ref{eqn:yieldFormula}), $11400\pm1500$.

\section{Results for the \Xb\ search}
\label{sec:results}

\subsection{Hypothesis tests}
\label{subsec:hyp}

A hypothesis test for the presence of an \Xb\ peak is performed
every $10\,\mev$ from $10\,\gev$ to $11\,\gev$, assuming a narrow state\footnote{
  Here, a narrow state refers to one whose natural width is much smaller than the experimental resolution.
} that has a differential cross section with a $(|y|,p_{\rm T})$ distribution similar
to that of the \upTw\ or \upTh, decaying according to three-body phase space.
The signal shape and bin splittings are treated as
described in Section~\ref{sec:fitting}. At each mass, a
simultaneous fit to the analysis bins is performed in a range
$m\pm8\sigma_{\text{ec}}$, where $\sigma_{\text{ec}}(m)$ is the width of the narrow signal component in the endcap: the
window varies from $\pm72\,\mev$ at $10\,\gev$ to $\pm 224\,\mev$ at $10.9\,\gev$;
near $11\,\gev$, $m-8\sigma_{\text{ec}}<m<11.2\,\gev$ is used.
When fitting near the \upTw\ and \upTh\ regions, the lineshapes for
these signal components are added to the background model,
with normalisations governed by Gaussian constraints to the values
obtained from the fits given in Section~\ref{sec:up2S3Sresults}.
In the immediate vicinity of the peaks,
$m_{\text{2S,3S}}\pm4\sigma_{\text{b}}$, no search is performed; the
width of the narrow signal component in the barrel, $\sigma_{\text{b}}$,
is $5.66\,\mev$ ($9.37\,\mev$) at the \upTw\ (\upTh) mass. This reduces
the analysis range to 10.05--10.31 and 10.40--$11.00\,\gev$.

\begin{figure}
  \centering
  \includegraphics[width=0.48\textwidth]{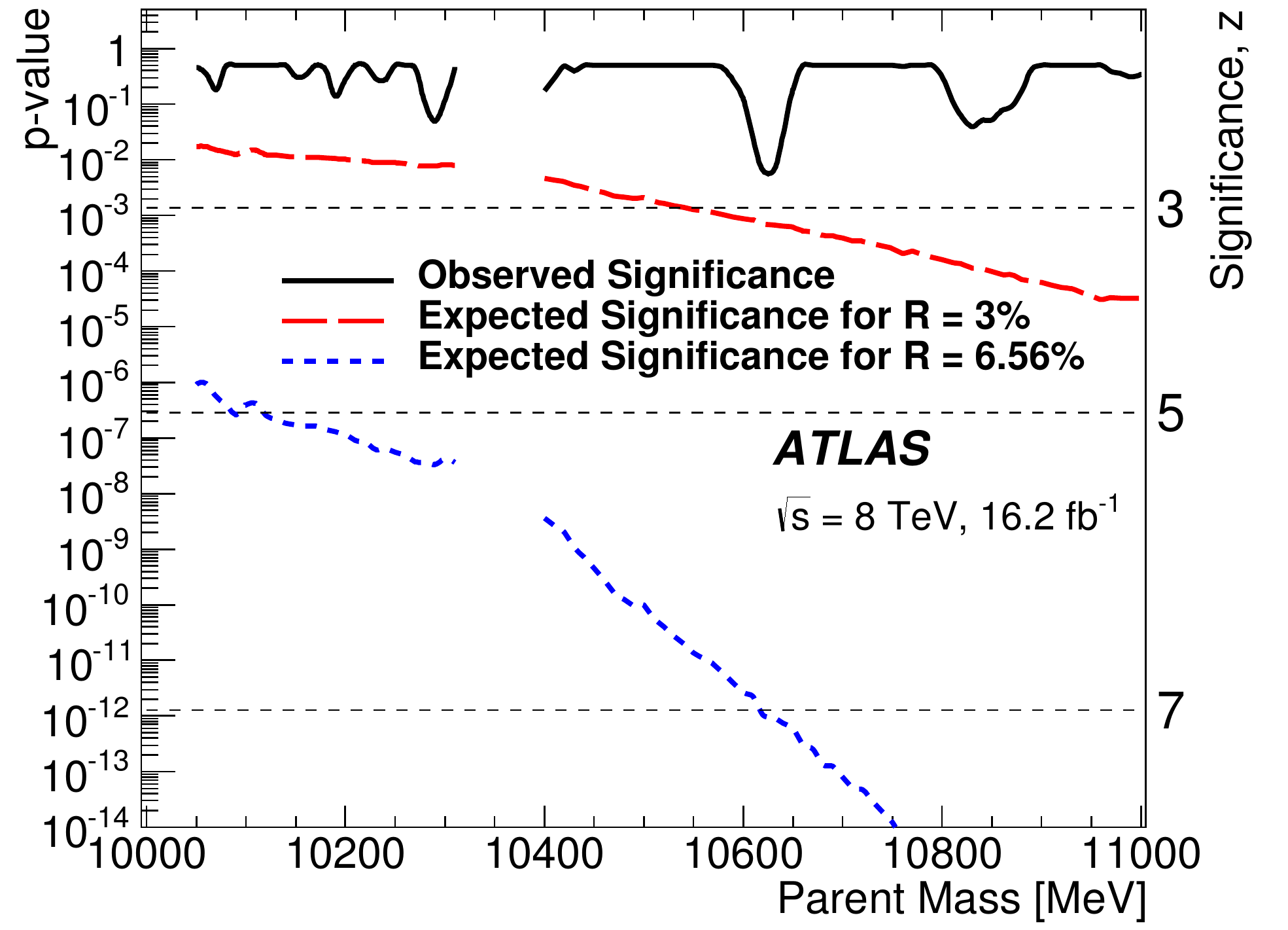}
  \caption{The solid curve shows the observed local $p$-value for the
	background-only hypothesis (left scale),
	and the corresponding significance, $z$, of a peak in \upp\
	(right scale), as a function of the mass of a hypothetical \Xb\
	parent state.
	Also shown are the expected values for the case of a signal with
	relative production
	rates $(\sigma\mathcal{B})/(\sigma\mathcal{B})_{\text{2S}}$
	of 3\% (red, long-dashed) and 6.56\% (blue, dashed curve).}
  \label{fig:upPiPiData2012_pvalueVsMass}
\end{figure}

At each mass, the $p$-value is extracted
using the asymptotic formula~\cite{asympForms} for the $q_{0}$ statistic,
a modification of the standard likelihood ratio (see Fig.~\ref{fig:upPiPiData2012_pvalueVsMass}).
No evidence for new states with local significance $z \geq 3$ is found.

The expected number of \Xb\ events can be written as
\begin{equation}
\label{eqn:R}
N = N_{\text{2S}}\cdot R \cdot \frac{\mathcal{A}}{\mathcal{A}_{\text{2S}}}\cdot \frac{\epsilon}{\epsilon_{\text{2S}}},
\end{equation}
where $R \equiv (\sigma\mathcal{B})/(\sigma\mathcal{B})_{\text{2S}}$,
the production rate relative to that of the \upTw.
The efficiency, $\epsilon$, as a function of mass is determined by
fitting a function $a + b/(1+e^{-c(m-d)})$ to the values from the simulated samples.
The acceptance increases with the mass of the parent state due to the increased
energy available to the pions and muons. Additionally, the measured production spectra
of the \up, \upTw, and \upTh\ states \cite{upProdXsec} are progressively harder in $p_{\rm T}$
with increasing mass. The acceptance for a hypothetical \Xb\ state of arbitrary mass is estimated here
by linear extrapolation of calculations performed at the \upTw\ and \upTh, using the measured production spectra of these states.
The ratio $(\mathcal{A}\epsilon)/(\mathcal{A}\epsilon)_{\text{2S}}$
rises from 1.1 at $m=10\,\gev$ to 7.5 at $m=11\,\gev$.

Using Eq.~(\ref{eqn:R}) and the formalism from Ref.~\cite{asympForms},
the expected significance for a relative production rate $R=6.56\%$
(the value of the analogous quantity for the $X(3872)$~\cite{cmsX3872})
is calculated as a function of mass, shown as
the dashed blue line in Fig.~\ref{fig:upPiPiData2012_pvalueVsMass};
it exceeds $5\sigma$ for $m \gtrsim 10.12\,\gev$.
Expectations for a weaker signal with $R=3\%$ are also shown (long-dashed red line).
Given the null result, upper limits are
calculated on R after modifying the fit to include systematic
uncertainties.

\subsection{Systematic uncertainties}
The upper limit calculation depends indirectly on signal and background
fitting parameters, including the fraction of the signal falling in each of
the analysis bins.
From Eq.~(\ref{eqn:R}), the upper limit on $R$ is proportional to
the inverse fitted \upTw\ yield, $N^{-1}_{\text{2S}}$,
and the ratios
$\mathcal{A}_{\text{2S}}/\mathcal{A}$ and $\epsilon_{\text{2S}}/\epsilon$.
For each source of systematic uncertainty, the impact on these factors is
quantified to find the maximum shift across the mass range.
These are then summed in quadrature and included in the fit
as Gaussian-constrained nuisance parameters.

The $X(3872) \to \pi^+\pi^-J/\psi$ dipion mass distribution
favours high mass~\cite{Abulencia:2005zc,Choi:2011fc};
for a potential hidden-beauty counterpart this distribution is unknown.
For $\psi(\text{2S}) \rightarrow \pi^+\pi^- J/\psi$ \cite{psi2SpipiJpsi},
and both \upTw~\cite{cleo2Sto1Spipi} and $\Upsilon(\text{4S}) \rightarrow \upp$ \cite{babar4Sto1S,belle4Sto1S},
the dipion mass distributions are concentrated near the upper boundary;
those for $Y(4260) \rightarrow \pi^+\pi^- J/\psi$ \cite{babarY4260}
and \upThupp\ \cite{cleoUpTransitions} are double-humped.
The results quoted here assume decay according to three-body phase space;
\upTw- and \upTh-like distributions
change the splitting functions by up to 35\%,
decrease the efficiency ratio by up to 17\%,
and produce modest changes in other parameters.

The next largest contribution is due to the linear extrapolation of the
acceptance between the \upTw\ and \upTh\ values. Alternative extrapolations
between the \up\ and \upTw, and between \up\ and \upTh, are also tried;
the greatest change in the acceptance ratio, 12\%,
is assigned as the uncertainty.

The parameters of the efficiency, the splitting functions,
and the widths of the narrow signal components $\sigma_{\text{b}}$ and $\sigma_{\text{ec}}$
as functions of mass, are varied by the uncertainties on
their fitted values; alternative functional forms are also tried.
In each case, the largest deviation is assigned as the systematic uncertainty.
The use of production weights (described in Section~\ref{sec:dataSimSamples}) relies on assumptions regarding
rapidity dependence,
and evolution from $\sqrt{s} = 7\,\tev$ to $8\,\tev$.
Removing these weights produces a ${\sim} 1\%$ change in efficiency ratio
(most of the differences cancel),
but changes the values of the splitting functions by up to 8\%.

Data versus simulation differences in the \upTw\ width parameters
in the barrel and endcap (1.9\% and 4.2\%, respectively) are incorporated as a source
of uncertainty, as is the statistical uncertainty on the averages used for
signal shape parameters $f$ and $r$ (0.5--1.4\%).
The background shape model is also altered, allowing a third-order term
comparable in size to typical values of the second-order terms.
Finally, uncertainties on $N_{\text{2S}}$ and the barrel/endcap scaling factor
are assigned based on uncertainties from the \upTw\ fits.

\subsection{Upper limit calculation}

\begin{figure}
  \centering
  \includegraphics[width=0.5\textwidth]{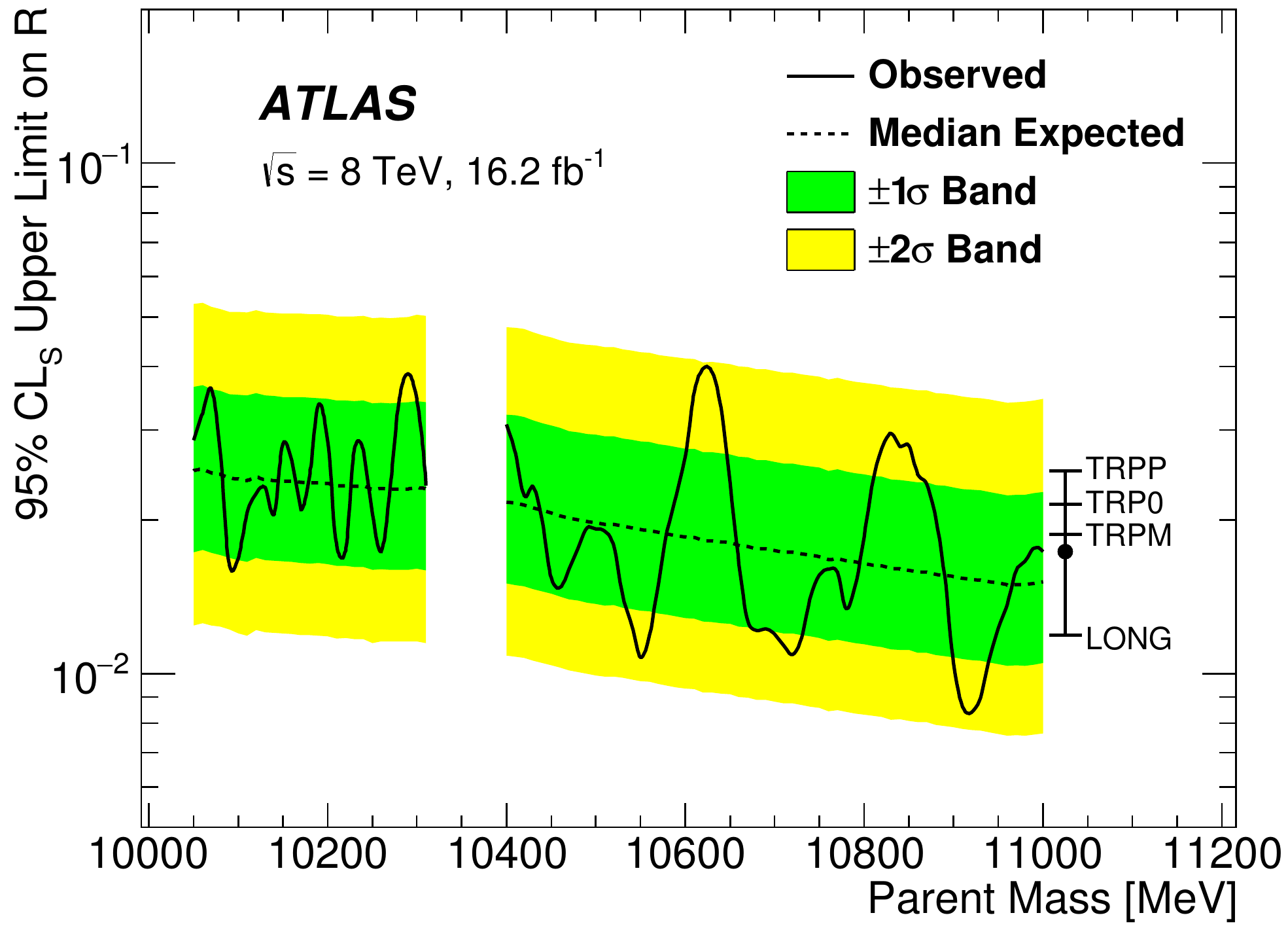}
  \caption{Observed 95$\%$ CL$_S$ upper limits (solid line)
	on the relative production rate
	$R=(\sigma\mathcal{B})/(\sigma\mathcal{B})_{\text{2S}}$
	of a hypothetical \Xb\ parent state decaying isotropically to \upp,
	as a function of mass.
	The median expectation (dashed)
	and the corresponding $\pm1\sigma$ and $\pm2\sigma$ bands
	(green and yellow respectively) are also shown.
	The bar on the right shows typical shifts under alternative
	\Xb\ spin-alignment scenarios, relative to the isotropic (`FLAT')
	case shown with the solid point.}
  \label{fig:upPiPiData2012_upperLimitVsMass}
\end{figure}

Upper limits are evaluated at the 95$\%$ confidence level using the CL$_S$
method by implementing asymptotic formulae for the $\tilde{q}_{\mu}$
statistic~\cite{asympForms}. The results
(Fig.~\ref{fig:upPiPiData2012_upperLimitVsMass}, solid line)
range between $R=0.8\%$ and 4.0\%. Median expected upper limits
assuming background only (dashed line),
and corresponding $\pm1\sigma$ and $\pm2\sigma$ bands are also shown.
These limits include the effect of systematic uncertainties: their inclusion
increased the observed limits by up to 13\% and inflated the $\pm1\sigma$
band by 9.5--25\%, depending on the \Xb\ mass.

As a check, upper limits are recalculated with modified fitting ranges
($m\pm7\sigma_{\text{ec}}$ and $m\pm9\sigma_{\text{ec}}$) and doubled bin widths
in the \upp\ mass distributions:
shifts are small compared to the $\pm1\sigma$ bands.
If an \upTw-like $m_{\pi^+\pi^-}$ distribution is assumed
(\emph{cf.}\ CMS~\cite{cmsAnalysis}),
expected upper limits increase: the fractional change
is $+17\%$ at $10.1\,\gev$, and ${\sim}{+}5\%$ for $m > 10.4\,\gev$.

These results exclude \Xb\ states with $R=6.56\%$
for masses 10.05--$10.31\,\gev$ and 10.40--$11.00\,\gev$.
The expected upper limits are more restrictive than those from CMS above
$m {\sim}10.1\,\gev$, and improve as a function of mass;
the discrimination in $(p_{\rm T}, \cos{\theta^{*}})$,
exploited by the binning method,
becomes increasingly important as mass increases.

If an \Xb\ state exists and lies within the range of masses
to which this analysis is sensitive,
its production cross section and/or its
branching fraction must be lower, relative
to the \upTw, than that of the $X(3872)$
relative to the $\psi(\text{2S})$.
There are arguments that the decay $\Xb\to\upp$ should be suppressed,
in the absence of the strong isospin-violating effects that are present for
$X(3872)\to\pi^+\pi^-J/\psi$~\cite{isospinXb,marekXb}.
In this case the \Xb\ would have more prominent decays to
$\pi^+\pi^-\chi_{b1}$, $\pi^+\pi^-\pi^0\up$, and other final states
which are relatively difficult to reconstruct.

All results to this point assume that any hypothetical \Xb\ production
is unpolarised.
Angular distributions of the $\Upsilon(\text{1S,2S,3S})$ states in $pp$
collisions are consistent
with unpolarised production~\cite{cmsUpsilonPolarisation},
but the \Xb\ spin-alignment is unknown and can have a strong impact
on the efficiency ratio, acceptance ratio, and bin splitting fractions.
Rather than including this as a systematic uncertainty,
upper limits are recalculated under
longitudinal (`LONG') and three transverse (`TRPP', `TRP0', `TRPM')
spin-alignment scenarios~\cite{jpsimumu}.
Shifts in the upper limits (either up or down)
depend only weakly on mass; the shift is smaller at large masses.
In Fig.~\ref{fig:upPiPiData2012_upperLimitVsMass} the effect of each hypothesis is
represented by a single number, chosen as the maximum difference in the median expected significance from the unpolarised (`FLAT') case.

\section{Results for the \upDJ\ triplet, $\Upsilon(10860)$, and $\Upsilon(11020)$}
\label{sec:results-other}

The search described above does not account for the closely spaced \upDJ\
triplet or the broad $\Upsilon(10860)$ and $\Upsilon(11020)$.
To fit for the \upDJ,
two extra peaks are added to the signal model.
CLEO~\cite{cleo1D} and BaBar~\cite{babar1D} have measured the \upDJTw\ mass,
with an average of $(10163.7\pm1.4)\,\mev$,
but the mass splitting within the triplet is unknown.
Averaging over several models~\cite{up1Dsplitting}
leads to triplet masses 10156, 10164, and $10170\,\mev$ (at $1\,\mev$ precision).
A fit is performed using these values,
assuming independent normalisations but common signal shapes
and bin splitting fractions. A significance of $z=0.12$ is found,
with fitted yields $-1000\pm3100$, $600\pm1800$, and $800\pm2300$ for $J=1$, 2, and 3.
Reasonable changes to the mass splittings do not appreciably increase the
significance, so there appears to be no evidence for \upDJ\ production.
Assuming that $J=2$ production dominates, or that the mass splitting is larger than
the experimental resolution, the upper limit on $R$ can be read from
Fig.~\ref{fig:upPiPiData2012_upperLimitVsMass}; combined with the measured
\upDJTwupp\ branching fraction~\cite{babar1D}, this yields an upper limit
on the relative cross section
$\sigma(pp \rightarrow \upDJTw)/\sigma(pp \rightarrow \upTw) \leq 0.55$.

The signal model for $\Upsilon(10860)$ and $\Upsilon(11020)$ is described
in Section \ref{sec:fitting}. Due to the large natural widths of these states,
the fitting range is extended to 10.498--11.198 GeV and the
background polynomial order increased to three. Significances $z=0.6$ and $z=0.3$
are found for $\Upsilon(10860)$ and $\Upsilon(11020)$,
for masses and widths fixed to world-average values~\cite{pdg}.
As these parameters have large uncertainties, the significance is also
calculated in a grid of $m\pm20$ MeV and $\Gamma\pm\Delta\Gamma$,
where $\Delta \Gamma$ is the uncertainty on the world-average width~\cite{pdg}.
The largest significance for the $\Upsilon(10860)$
is $z=1.1$ at $m=10856$ MeV and $\Gamma=55$ MeV.
For the $\Upsilon(11020)$, the largest significance is $z=0.6$ at $m=11039$ MeV and $\Gamma=95$ MeV.
Thus, no evidence for $\Upsilon(10860)$ or $\Upsilon(11020)$ production is found.

\section{Conclusions}

A search for a hidden-beauty analogue of the $X(3872)$ is conducted
by reconstructing $\upp (\rightarrow \mu^+\mu^-)$ events in $16.2\,\invfb$
of $pp$ collision data recorded at $\sqrt{s}=8\,\tev$ by ATLAS at the LHC.
To optimise the sensitivity of the search, the analysis is performed in eight bins of
rapidity, transverse momentum, and the angle (in the rest frame of the parent state) between the dipion system
and the laboratory-frame momentum of the parent.
At each mass, the presence of a signal is tested
by performing simultaneous fits to the nearby \upp\ mass spectrum in
these bins; no evidence for new narrow states is found for masses 10.05--$10.31\,\gev$
and 10.40--$11.00\,\gev$. Upper limits are also set on the ratio
$R = [ \sigma(pp \to \Xb) \mathcal{B}(\Xb\to\pi^+\pi^-\Upsilon(\text{1S})) ] / [ \sigma(pp \to \Upsilon(\text{2S})) \mathcal{B}(\Upsilon(\text{2S}) \to \pi^+\pi^-\Upsilon(\text{1S})) ]$,
with results ranging from 0.8\% to 4.0\% depending on the \Xb\ mass.
The analogous ratio for the $X(3872)$ is 6.56\%: a value this large
is excluded for all \Xb\ masses considered.
Separate fits to the \upDJ\ triplet,
$\Upsilon(10860)$, and $\Upsilon(11020)$
also reveal no significant signals, and a CL$_S$ upper limit of 0.55 is
set on $\sigma(pp \rightarrow \upDJTw)/\sigma(pp \rightarrow \upTw)$.

\section{Acknowledgements}
We thank CERN for the very successful operation of the LHC, as well as the
support staff from our institutions without whom ATLAS could not be
operated efficiently.

We acknowledge the support of ANPCyT, Argentina; YerPhI, Armenia; ARC,
Australia; BMWFW and FWF, Austria; ANAS, Azerbaijan; SSTC, Belarus; CNPq and FAPESP,
Brazil; NSERC, NRC and CFI, Canada; CERN; CONICYT, Chile; CAS, MOST and NSFC,
China; COLCIENCIAS, Colombia; MSMT CR, MPO CR and VSC CR, Czech Republic;
DNRF, DNSRC and Lundbeck Foundation, Denmark; EPLANET, ERC and NSRF, European Union;
IN2P3-CNRS, CEA-DSM/IRFU, France; GNSF, Georgia; BMBF, DFG, HGF, MPG and AvH
Foundation, Germany; GSRT and NSRF, Greece; ISF, MINERVA, GIF, I-CORE and Benoziyo Center,
Israel; INFN, Italy; MEXT and JSPS, Japan; CNRST, Morocco; FOM and NWO,
Netherlands; BRF and RCN, Norway; MNiSW and NCN, Poland; GRICES and FCT, Portugal; MNE/IFA, Romania; MES of Russia and ROSATOM, Russian Federation; JINR; MSTD,
Serbia; MSSR, Slovakia; ARRS and MIZ\v{S}, Slovenia; DST/NRF, South Africa;
MINECO, Spain; SRC and Wallenberg Foundation, Sweden; SER, SNSF and Cantons of
Bern and Geneva, Switzerland; NSC, Taiwan; TAEK, Turkey; STFC, the Royal
Society and Leverhulme Trust, United Kingdom; DOE and NSF, United States of
America.

The crucial computing support from all WLCG partners is acknowledged
gratefully, in particular from CERN and the ATLAS Tier-1 facilities at
TRIUMF (Canada), NDGF (Denmark, Norway, Sweden), CC-IN2P3 (France),
KIT/GridKA (Germany), INFN-CNAF (Italy), NL-T1 (Netherlands), PIC (Spain),
ASGC (Taiwan), RAL (UK) and BNL (USA) and in the Tier-2 facilities
worldwide.

\section*{References}
\bibliography{references-XbPaper.bib}
\bibliographystyle{elsarticle-num}

\onecolumn
\clearpage
\begin{flushleft}
{\Large The ATLAS Collaboration}

\bigskip

G.~Aad$^{\rm 84}$,
B.~Abbott$^{\rm 112}$,
J.~Abdallah$^{\rm 152}$,
S.~Abdel~Khalek$^{\rm 116}$,
O.~Abdinov$^{\rm 11}$,
R.~Aben$^{\rm 106}$,
B.~Abi$^{\rm 113}$,
M.~Abolins$^{\rm 89}$,
O.S.~AbouZeid$^{\rm 159}$,
H.~Abramowicz$^{\rm 154}$
% ...
\bigskip
\\
% ...
$^{11}$ Institute of Physics, Azerbaijan Academy of Sciences, Baku, Azerbaijan\\
% ...
$^{84}$ CPPM, Aix-Marseille Universit{\'e} and CNRS/IN2P3, Marseille, France\\
% ...
$^{89}$ Department of Physics and Astronomy, Michigan State University, East Lansing MI, United States of America\\
% ...
$^{106}$ Nikhef National Institute for Subatomic Physics and University of Amsterdam, Amsterdam, Netherlands\\
% ...
$^{112}$ Homer L. Dodge Department of Physics and Astronomy, University of Oklahoma, Norman OK, United States of America\\
$^{113}$ Department of Physics, Oklahoma State University, Stillwater OK, United States of America\\
% ...
$^{116}$ LAL, Universit{\'e} Paris-Sud and CNRS/IN2P3, Orsay, France\\
% ...
$^{152}$ Institute of Physics, Academia Sinica, Taipei, Taiwan\\
% ...
$^{154}$ Raymond and Beverly Sackler School of Physics and Astronomy, Tel Aviv University, Tel Aviv, Israel\\
% ...
$^{159}$ Department of Physics, University of Toronto, Toronto ON, Canada\\
% ...
\end{flushleft}

\end{document}